\documentclass[11pt]{article}

\usepackage{amsfonts,amsmath,amssymb}
\usepackage{enumerate}
\usepackage{hyperref}
\usepackage{bbm}
\usepackage[all]{xy}
\usepackage{graphicx}
\usepackage{bm}

\usepackage{upgreek}

\usepackage{cite}

\usepackage{booktabs}

\newcommand{\be}{\begin{equation}}
\newcommand{\ee}{\end{equation}}

\newcommand{\bea}{\begin{eqnarray}}
\newcommand{\eea}{\end{eqnarray}}

\def\ie{\begin{equation}\begin{aligned}}
\def\fe{\end{aligned}\end{equation}}

\newcommand{\bes}{\begin{subequations}}
\newcommand{\ees}{\end{subequations}}

\def\sst#1{{\scriptscriptstyle #1}}

\def\0{{\sst{(0)}}}
\def\1{{\sst{(1)}}}
\def\2{{\sst{(2)}}}
\def\3{{\sst{(3)}}}
\def\4{{\sst{(4)}}}
\def\5{{\sst{(5)}}}
\def\6{{\sst{(6)}}}
\def\7{{\sst{(7)}}}
\def\8{{\sst{(8)}}}

\newcommand{\ext}{\textrm{ext}}

\def\th{\theta}

\def\hr{\hat{r}}

\def\d{\delta}

\newcommand{\non}{\nonumber}
\def\half{{1\over 2}}

\usepackage{multirow}
\usepackage{rotating}

\allowdisplaybreaks

\begin{document}

\makeatletter
\renewcommand{\theequation}{\thesection.\arabic{equation}}
\@addtoreset{equation}{section}
\makeatother

\begin{titlepage}

\begin{flushright}
\end{flushright}

\vspace{55pt}

   \begin{center}
   \baselineskip=16pt
   \begin{Large}\textbf{
Logarithmic corrections to black hole entropy \\[10pt] from Kerr/CFT}
   \end{Large}

\vspace{40pt}

{Abhishek Pathak$^1$,  Achilleas P. Porfyriadis$^{1,2}$,  Andrew Strominger$^1$ and Oscar Varela$^{1,3,4}$}

\vspace{25pt}
		
\begin{small}
	
	$^1$ {\it Center for the Fundamental Laws of Nature,
		Harvard University, Cambridge, MA, USA } \\

	\vspace{5pt}
	
	{\it $^{2}$ Department of Physics, UCSB, Santa Barbara, CA, USA  } \\

	\vspace{5pt}
	
	{\it $^{3}$ Max-Planck-Institut f\"ur Gravitationsphysik (Albert-Einstein-Institut), Potsdam, Germany.  } \\

	\vspace{5pt}
	
	{\it $^{4}$ Department of Physics, Utah State University, Logan, UT, USA.} 
	
	\vspace{30pt}

\end{small}

\vskip 50pt

\end{center}
%
%
\begin{center}
\textbf{Abstract}
\end{center}

\begin{quote}It has been shown by A. Sen that logarithmic corrections to the black hole area-entropy law are entirely determined macroscopically from the massless particle spectrum. They therefore serve as powerful consistency checks on any proposed enumeration of quantum black hole microstates. 
Sen's results include a macroscopic computation of the logarithmic corrections for a five-dimensional near extremal Kerr-Newman black hole.
Here we compute these corrections microscopically using a stringy embedding of the Kerr/CFT correspondence  and  find perfect agreement. 
\end{quote}

\end{titlepage}

\tableofcontents


\section{Introduction}
 The Bekenstein-Hawking area-entropy law universally applies to any self-consistent quantum theory of gravity. 
 Efforts to understand how the former constrains the latter have led to a wealth of insights. 
 
 Five years ago Sen et al. \cite{Banerjee:2010qc,Banerjee:2011jp,Sen:2011ba,Sen:2012cj,Bhattacharyya:2012wz,Sen:2012dw} pointed out that the leading corrections to this law, which are of order $\log A$ (where $A$ is the area) are also universal
 in that they depend only on the massless spectrum of particles and are insensitive to the UV completion of the theory. The basic reason for this is that the effects of a particle  of mass $m$ can be accounted for by integrating it out, which generates local higher derivative  terms in the effective action. These lead to corrections to the entropy which are suppressed by inverse powers of $m^2A$ and cannot give $\log A$ terms.
 
 The macroscopically computed logarithms serve as a litmus test for any proposed enumeration of quantum black hole microstates which is more refined than the test provided by the area law. Sen extensively analyzed a number of stringy examples all of which passed the test with flying colors \cite{Sen:2014aja}, and further noted that the macroscopic computation does not match the loop gravity result. He also posed a matching of the logarithms as a challenge for Kerr/CFT \cite{Guica:2008mu,Bredberg:2009pv,Bredberg:2011hp,Compere:2012jk}. 
 
In this paper we show that the logarithms indeed match for one microscopic realization \cite{Guica:2010ej,Song:2011sr} of Kerr/CFT obtained by 
embedding a certain near-extremal five-dimensional Kerr-Newman black hole into a string compactification.\footnote{The near-extremal regime sidesteps subtleties with logarithmic corrections at extremality.} The microscopic dual is a two-dimensional field theory defined as the IR fixed point of the worldvolume field theory on a  certain brane configuration with scaled fluxes. This IR limit is certainly nontrivial but is not a conventional 2D CFT. Its properties are incompletely understood and have been studied in a variety of approaches: see $e.g.$ \cite{Hofman:2011zj,ElShowk:2011cm,Bena:2012wc,Azeyanagi:2012zd,Detournay:2012pc,Detournay:2012dz,Compere:2014bia}.

The example of Kerr/CFT we chose is the simplest possible case.  In the simplest examples considered by Sen, the macro-micro match has a somewhat trivial flavor: all logs vanish in a certain thermodynamic ensemble, with logs generated on both sides of the match by Legendre transforms to a different ensemble. However, later examples become quite intricate and provide compelling tests of a variety of stringy constructions. In the five-dimensional Kerr/CFT match herein, all logs vanish in a certain thermodynamic ensemble, and the character of the match is as in the simplest of Sen's examples. Non-trivial aspects of our match reside in its reliance on the values of the Kerr/CFT  central charges and Kac-Moody levels which enter the microscopic computation. In particular, we find that a necessary non-vanishing level of the current dual to the electric field is provided by a Chern-Simons term which is crucially present in the effective action obtained from string theory. While it is reassuring that the simplest case works, perhaps more challenging matches such as the extremal 4D case  may eventually provide 
more refined and compelling tests of Kerr/CFT. 

This paper is organized as follows. In section \ref{blackhole} we describe the black hole solution, take its near horizon limit, and determine the corresponding quantum mixed state in the CFT. In section \ref{logcorrections} we begin by stating the result of \cite{Sen:2012dw} for the logarithmic corrections to the Bekenstein-Hawking entropy and then proceed to compute them microscopically in the dual theory for two different enhancements of the global symmetries. We match the Bekenstein-Hawking entropy of the near extremal black hole to first order in the Hawking temperature with the Cardy formula and, using the result of Appendix \ref{App ASG} for the gauge Kac-Moody level, we show that the logarithmic corrections also agree. 

Previous work on logarithmic corrections to black hole entropy includes \cite{Solodukhin:1994yz,Solodukhin:1994st,Fursaev:1994te,Mann:1997hm,Carlip:2000nv,Ferrara:2011qf,Keeler:2014bra,Chowdhury:2014lza,Larsen:2014bqa,Charles:2015eha,Belin:2016knb}.

\section{The five-dimensional Kerr-Newman black hole}\label{blackhole}

We consider a charged and rotating black hole solution of five-dimensional Einstein gravity minimally coupled to a gauge field. The dynamics of the latter is specified by the Yang-Mills-Chern-Simons Lagrangian, so that the complete action is,\footnote{This coincides with the bosonic sector of minimal supergravity in five dimensions.} 
\be \label{5daction} 
S_5= \frac{1}{4\pi^2} \int d^5 x \left(\sqrt{-g} \left (R- \frac{3}{4}F^2\right)+ \frac{1}{4}
 \epsilon^{abcde}A_aF_{bc}F_{de}\right)\,.
\ee
Specifically, we are interested in the following Kerr-Newman black hole solution to (\ref{5daction}) considered in \cite{Guica:2010ej},
\bea
\label{5dmet}
ds_5^2 & = & - \frac{(a^2+\hr^2)(a^2+\hr^2-M_0) }{\Sigma^2}\, d\hat{t}^2 + 
\Sigma \left(\frac{\hr^2 d\hr^2}{f^2-M_0 \hr^2}+ \frac{d\th^2}{4}\right) - \frac{M_0 F}{\Sigma^2} \,
 (d\hat{\psi}+\cos\th\, d\hat{\phi})\, d\hat{t} \non \\
&& + \frac{\Sigma}{4} (d\hat{\psi}^2+d\hat{\phi}^2+2\cos\th \, d\hat{\psi}\, d\hat{\phi}) 
 + \frac{a^2 M_0 B}{4\Sigma^2}\, (d\hat{\psi}+\cos\th \,d\hat{\phi})^2 \; ,  \\[12pt]
\label{5dgaugefield}
 A & = & \frac{M_0  \sinh 2\delta }{2 \Sigma} \left( d\hat{t} - \half a e^\delta (d\hat{\psi} + \cos\th \, d\hat{\phi}) \right) \; ,
\eea
where we have defined the quantities
\bea
& B= a^2+\hr^2 -2 M_0 s^3 c^3 - M_0 s^4(2 s^2+3)\;, \qquad F= a(\hr^2+a^2) (c^3+s^3) - a M_0 s^3\,, \nonumber \\[4pt]
& \Sigma =  \hat{r}^2 + a^2 + M_0 s^2  \ , \qquad f = \hat{r}^2 + a^2 \,,
\eea
and $s \equiv \sinh \delta\,,c \equiv \cosh \delta$. 
The geometry depends on three independent parameters $(a, M_0, \delta)$ and the physical quantities of the black hole, {\it i.e.}, its mass, angular momentum and electric charge, are given in terms of those parameters by
\be \label{BHcharges}
M = {3M_0\over 2} \cosh 2\delta \; , \qquad 
J_L  = a M_0\,(c^3 + s^3 ) \; , \qquad 
Q  = M_0sc \; . 
\ee
In five dimensions, it is possible to have a second angular momentum, $J_R$, but we set $J_R=0$. Note that the $SU(2)_L$ angle is identified $\hat{\psi}\sim\hat{\psi}+4\pi$.

This black hole displays inner and outer horizons located at
\begin{equation} \label{horizons}
r_\pm^2 = \tfrac{1}{2} (M_0-2a^2) \pm \tfrac{1}{2} \sqrt{M_0(M_0 - 4a^2)} \; . 
\end{equation}
At the (outer) horizon, the angular velocities are
\be \label{BHangvel}
\Omega_L \equiv  \Omega_{\hat \psi} = \frac{4a}{M_0} \, \frac{1}{ (c^3- s^3)+(c^3+s^3) \sqrt{1- 4 a^2 / M_0} }  \;, \;\;\;\;\; \Omega_R \equiv  \Omega_{\hat \phi } =0 \; , 
\ee
and the electric potential is
\be \label{BHElecPot}
\Phi = \frac{c^2 s -s^2 c + (c^2s+s^2c) \sqrt{1- 4 a^2 / M_0}}{c^3-s^3 + (c^3+s^3)\sqrt{1- 4 a^2 / M_0} } \; . 
\ee
Finally, the Hawking temperature is given by
\begin{eqnarray} \label{THawking}
T_H = \frac{1}{\pi \sqrt{M_0}} \, \frac{ \sqrt{1- 4 a^2 / M_0}}{c^3-s^3 + (c^3+s^3)\sqrt{1- 4 a^2 / M_0} } \; , 
\end{eqnarray}
and the Bekenstein-Hawking entropy is 
\begin{eqnarray} \label{BHEntropy}
S_{BH} = \pi \sqrt{2} \, M_0 \, \sqrt{ (c^6+s^6) M_0  -2 (c^3+s^3)^2 a^2  +  (c^4+c^2s^2+ s^4) \sqrt{M_0 (M_0 - 4a^2)}} \, .
\end{eqnarray}

The black hole approaches extremality in the limit $M_0 \to 4a^2$. In this limit, the two horizons (\ref{horizons}) coalesce at $r_+ = a$ and the Hawking temperature (\ref{THawking}) vanishes. The charges (\ref{BHcharges}) become
\be \label{BHchargesExt}
M_{\textrm{ext}} = 6a^2  \cosh 2\delta \; , \qquad 
J_{L  \, \textrm{ext}}  = 4a^3 \,(c^3 + s^3 ) \; , \qquad 
Q_{\textrm{ext}}  = 4a^2 sc \; , 
\ee
and the angular velocity (\ref{BHangvel}) and electric potential (\ref{BHElecPot}) become
\be \label{BHangvelext}
\Omega_{L \, \ext} = \frac{1}{a(c^3-s^3)} \;, \qquad 
\Phi_{\textrm{ext}} = \frac{c^2 s -s^2 c}{c^3-s^3 } \; . 
\ee
At extremality, the Bekenstein-Hawking  entropy (\ref{BHEntropy}) reduces to
\begin{eqnarray} \label{BHEntropyExt}
S_{BH \, \ext} = 8 \pi a^3 (c^3 - s^3) \; .
\end{eqnarray}
In this paper we are interested in the near-extreme case so we introduce a small parameter $\hat{\kappa}$ that measures the deviation from extremality and write $M_0 = 4a^2 + a^2 \hat{\kappa}^2$. Substituting this into (\ref{BHEntropy}) and keeping terms up to linear order in $\hat{\kappa}$, the near extremal entropy is
\begin{eqnarray} \label{BHEntropyNearExt}
S_{BH \, \textrm{near ext}}  &=& 8 \pi a^3 (c^3 - s^3)  + 4 \pi a^3 (c^3 + s^3) \, \hat{\kappa}^2 + {\cal O} ( \hat{\kappa}^2 )  \nonumber \\[10pt]
&=&   \frac{\pi^2}{3}  \, (6 J_{L \, \ext} )  \left( \frac{1}{\pi} \frac{c^3-s^3} {c^3+s^3}  + \frac{\hat{\kappa} }{2\pi} \right)   + {\cal O} ( \hat{\kappa}^2 ) \; .
\end{eqnarray}
%

\subsection{Near horizon, near extremal limit}

Consider the coordinate transformation
\begin{eqnarray} \label{Coordchange}
t = \tfrac{1}{2} \,  \epsilon \,  \Omega_{L \, \ext} \, \hat{t} \;  , \qquad
 r = \frac{\hat{r}^2 - r_+^2}{\epsilon \, r_+^2} \; , \qquad 
\psi =  \hat{\psi} -\Omega_{L \, \ext} \hat{t} \; , \qquad 
\phi =  \hat{\phi} \; .
\end{eqnarray}
Here, $r_+$ is the location of the outer horizon given in (\ref{horizons}) and  $\Omega_{L \, \ext}$ is the extremal angular velocity (\ref{BHangvelext}). Making this coordinate transformation in the five-dimensional geometry (\ref{5dmet}), (\ref{5dgaugefield}), with $M_0$ fixed to its extremal value, $M_0 = 4 a^2$, and letting $\epsilon \rightarrow 0$, one obtains the extremal near horizon geometry given in \cite{Guica:2010ej}. 

Here, we are interested in reaching the near horizon geometry of the black hole close, but not exactly at, extremality. This is the analog of the so-called near-NHEK limit for 4D Kerr considered in  \cite{Bredberg:2009pv}. In order to do this, we still make the coordinate transformation (\ref{Coordchange}), but now parametrize deviations from extremality with a parameter $\kappa$ defined by 
\begin{equation} \label{M0nearext}
M_0 = 4a^2 + a^2 \epsilon^2 \kappa^2 \; .
\end{equation}
Then the metric (\ref{5dmet}) gives rise to
\bea
\label{5dmetnearext}
ds_5^2 &=& \frac{M_\ext}{12} \left[-r(r+2\kappa) dt^2 + \frac{dr^2}{r(r+2\kappa)} +d\theta^2 +\sin^2\theta d\phi^2 \right.
\nonumber \\
&& \qquad \quad \left. +\frac{27 J_{L \, \ext}^2}{M_\ext^3} \big(  \pi T_{L} ( d\psi +  \cos \th\, d\phi )  +(r + \kappa) dt \big)^2 \right]
\eea
in the $\epsilon \rightarrow 0$ limit. Here, we have defined 
\begin{equation} \label{TLpreview}
T_{L} \equiv \frac{1}{\pi} \,  \frac{c^3-s^3}{c^3+s^3} \; .
\end{equation}
This notation will be clarified in the next subsection. 
The location of the horizon in (\ref{5dmetnearext}) is at $r=0$ and the associated surface gravity is $\kappa$. We denote the corresponding Hawking temperature by 
\begin{equation} \label{TRpreview}
T_R \equiv \frac{\kappa}{2\pi} \; .
\end{equation}

When we identify $\kappa$ with the parameter $\hat{\kappa}$ introduced in (\ref{BHEntropyNearExt}), the metric \eqref{5dmetnearext} corresponds to the near horizon geometry of the black hole \eqref{5dmet} close to extremality in the following complementary sense as well. Making the coordinate transformation \eqref{Coordchange} with $\epsilon=1$ and expanding the metric components in \eqref{5dmet} to leading order in $r\sim\hat{\kappa}\ll 1$ we obtain \eqref{5dmetnearext} with $\kappa=\hat{\kappa}$. In the rest of the paper we make this identification throughout.

The gauge field corresponding to the near horizon, near extremal geometry is obtained by accompanying the coordinate transformation (\ref{Coordchange}) with the gauge transformation
\begin{equation} \label{gaugetrans}
A \,  \rightarrow \, A -d \Lambda \; , \qquad \textrm{with} \quad \Lambda \equiv  \Phi_\ext \, \hat{t} \,.
\end{equation}
Then the gauge field (\ref{gaugetrans}) becomes
\be \label{aeq}
A=-\half a e^\d \tanh 2 \d \left( d\psi  + \cos\th d\phi + e^{-2 \d} (r + \kappa) dt \right) 
\ee
in the $\epsilon \rightarrow 0$ limit.

\subsection{Frolov-Thorne temperatures}

We now move on to compute the Frolov-Thorne temperatures corresponding to the near-extremal Kerr-Newman black hole, by adapting the strategy of \cite{Bredberg:2009pv} to our present context. Consider a scalar field
\begin{eqnarray} \label{scalarfield}
\varphi = e^{-i \omega \hat t  + i m \hat \psi } \,  \hat R(\hat r) \, S( \theta) \, T(\hat{\phi}) \;
\end{eqnarray}
on the the black hole background (\ref{5dmet}), with charge $q$ under the gauge field (\ref{5dgaugefield}). Zooming into the near horizon region requires performing the coordinate transformation (\ref{Coordchange}) combined with the gauge transformation (\ref{gaugetrans}). The charged scalar (\ref{scalarfield}) thus becomes
\begin{eqnarray} \label{scalarfieldnearhor}
\varphi = e^{i q \Lambda  } e^{-i n_R t  + i n_L  \psi } \,  R (r) \, S( \theta) \, T (\phi)  \;
\end{eqnarray}
with
\begin{eqnarray} \label{identifications}
\Lambda = \frac{2 \Phi_\ext}{\epsilon \,  \Omega_{L \, \ext}} \, t \; , \quad 
m = n_L \; , \quad 
\omega  = \frac{1}{2}  \epsilon \, \Omega_{L \, \ext} \Big( n_R + \frac{2}{\epsilon} \, n_L  - \frac{2 q   \Phi_\ext}{\epsilon \,  \Omega_{L \, \ext}} \Big) \; .
\end{eqnarray}
Now, the scalar field is in a mixed quantum state whose density matrix has eigenvalues given by the Boltzmann factor $e^{- \frac{1}{T_H} (\omega - m \Omega_L + q \Phi)}$, where $T_H$ is the Hawking temperature (\ref{THawking}). Identifying
\begin{equation}
e^{- \frac{1}{T_H} (\omega - m \Omega_L + q \Phi)} =  e^{-\frac{2n_L}{T_L}-\frac{n_R}{T_R}-\frac{q}{T_Q}}
\end{equation}
and using (\ref{identifications}) we find the following Frolov-Thorne temperatures:
\begin{eqnarray}
\label{TL}
T_R & =& \frac{2}{\epsilon \, \Omega_{L\textrm{ext}}} T_H = \frac{2a}{\epsilon \pi \sqrt{M_0}} \, \frac{ (c^3-s^3) \sqrt{1-4a^2/M_0}}{c^3-s^3 + (c^3+s^3) \sqrt{1-4a^2/M_0} } \; , \\[5pt]
\label{TR}
T_L & = & -\frac{2}{  \Omega_L- \Omega_{L\textrm{ext}} } \, T_H =  \frac{2a}{ \pi \sqrt{M_0}} \, \frac{ (c^3-s^3) }{c^3 +s^3 + (c^3-s^3) \sqrt{1-4a^2/M_0} }  \; , \\[5pt] 
\label{TQ}
T_Q & =&  \frac{1}{  \Phi- \Phi_{\textrm{ext}} } \, T_H = \frac{1}{2\pi \sqrt{M_0} } \frac{c^3-s^3}{s^2c^2} \; .
\end{eqnarray}
Near extremality, $M_0$ is given by (\ref{M0nearext}) and (\ref{TL})--(\ref{TQ}) become, in the $\epsilon \rightarrow 0 $ limit,
\begin{eqnarray}
\label{Tnearext}
T_R = \frac{\kappa}{2\pi}  \; , \qquad 
\label{TRnearext}
T_L = \frac{1}{\pi} \frac{c^3-s^3} {c^3+s^3}  \; , \qquad
\label{TQnearext}
T_Q = \frac{1}{4\pi a} \frac{c^3-s^3}{s^2c^2}    \; .
\end{eqnarray}
Recall that both $T_R$ and $T_L$ have already appeared in our discussion: the former as the Hawking temperature (\ref{TRpreview}) of the near-horizon, near-extremal metric (\ref{5dmetnearext}) and the latter as a parameter, (\ref{TLpreview}), in that metric. The present analysis elucidates the names given previously to those quantities.

\section{Logarithmic correction to entropy}\label{logcorrections}
The logarithmic correction to the microcanonical entropy of a non-extremal, rotating charged black hole in general spacetime dimension $D$ has been computed by Sen in \cite{Sen:2012dw}. His result applies to the near extremal black hole considered in this paper, which has a small but non zero Hawking temperature. Equation (1.1) in \cite{Sen:2012dw} for the correction to the microcanonical entropy reads
\be\label{smcsen}
S_{mc}\left(M, \vec{J}, \vec{Q}\right) = S_{BH}\left(M,\vec{J}, \vec{Q}\right) + \log a \left(C_{local} -\frac{D-4}{2} - \frac{D-2}{2}N_C - \frac{D-4}{2} n_V\right) 
\ee
where $N_C = \left[ \frac{D-1}{2} \right]$ is the number of Cartan generators of the spatial rotation group and $n_V$ is the number of vector fields in the theory. $C_{local}$ arises from one loop determinants of massless fields fluctuating in the black hole background and vanishes in odd dimensions. The remaining contribution in \eqref{smcsen} comes from zero modes and Legendre transforms. Plugging $D = 5$, $N_C=2$ and $n_V = 1$ into  \eqref{smcsen} we have
\be\label{DeltaSmacro}
S_{mc} = S_{BH} - 4 \log a \; , 
\ee
with, for the case at hand, $S_{BH}$ given by (\ref{BHEntropy}).

\subsection{Microscopic computation} \label{microscopiccomputation}

We now change gears and compute the logarithmic correction to the entropy of the microscopic theory dual to the Kerr-Newman black hole. In \cite{Guica:2010ej} this solution was embedded into string theory and the microscopic dual thereby shown to be the  infrared fixed point of a 1+1 field theory  living on the brane intersection.  This fixed point is a possibly non-local deformation of an ordinary 1+1 conformal field theory which preserves at least one infinite-dimensional conformal symmetry. While the string theoretic construction implies the existence of the fixed point theory, it exhibits a new kind of 1+1 $D$ critical behavior and is only partially understood. The near horizon geometry \eqref{5dmetnearext} has a $SL(2,R)_R \times U(1)_L$ isometry subgroup coming from the isometries of the  $AdS_2$ submanifold  and the unbroken $U(1)_L\subset SU(2)_L$ rotation isometry respectively.  
Various infinite-dimensional enhancements of this global isometry, involving different boundary conditions, have been extensively considered in the literature, and may be relevant in different circumstances or for different computations.  See \cite{Compere:2014bia} for a recent discussion. We consider two of them which turn out to both give the same log corrections.\footnote{Had they been different, the matching of logarithmic corrections would have singled one out.}

\subsubsection{$Vir_R \times Vir_L$}

In this subsection we consider a CFT in which the global symmetries are enhanced as
\be\label{enhancement2}
SL(2,R)_R \times U(1)_L \rightarrow Vir_R \times Vir_L \; ,
\ee
where $Vir_L$  and $Vir_R$ are left and right moving Virasoro algebras  with generators $L_n$ and $\bar{L}_n$ respectively. $L_0$ generates $\psi$ rotations  and $\bar{L}_0$ generates $AdS_2$ time translations.

We put the CFT on a circle along $\psi-t$ and consider the ensemble
\be
Z(\tau,\bar{\tau})=\textrm{Tr}\,e^{2\pi i \tau L_0-2\pi i\bar{\tau}\bar{L}_0}\,.
\ee
We assume that 
\be\label{tau vs beta cft}
4\pi \tau=\beta_L-\beta_R+i(\beta_L+\beta_R)
\ee
and $4\pi \bar{\tau}=\beta_L-\beta_R-i(\beta_L+\beta_R)$. 
Standard modular invariance of this partition function is $Z(\tau,\bar{\tau})=Z(-1/\tau,-1/\bar{\tau})$.
The microscopic dual to the Kerr-Newman black hole we are considering in this paper has an additional $SU(2)\times U(1)$ global symmetry, corresponding to the $SU(2)$ rotation isometry and the $U(1)$ gauge symmetry. Turning on the associated chemical potentials, the partition function becomes
\be\label{Z with chem pot}
Z(\tau,\bar{\tau},\vec{\mu})=\textrm{Tr}\,e^{2\pi i \tau L_0-2\pi i\bar{\tau}\bar{L}_0+2\pi i\mu_i P^i}
\ee
and it obeys the modular transformation rule
\be\label{mod transfn}
Z(\tau, \bar{\tau}, \vec{\mu}) = e^{- \frac{2\pi i \mu^2}{ \tau}} Z \left (-\frac{1}{\tau}, -\frac{1}{\bar{\tau}}, \frac{\vec{\mu}}{\tau}  \right)\,. 
\ee
Here $\mu_i$ are left chemical potentials associated with the left moving conserved charges $P^i$ and $ \mu^2 \equiv  \mu_i \mu_j k^{ij} $ with $k^{ij}$ the matrix of Kac-Moody levels of the left moving currents. In our case $i,j$ run from $1$ to $2$ but, for the sake of generality, we temporarily assume they run from $1$ to $n$. 
This partition function is related to the density of states, $\rho$, at high temperatures by
\be\label{Z vs d}
Z(\tau, \bar{\tau}, \vec{\mu}) = \int d E_L~ d E_R ~ d^{n}p ~ \rho(E_L, E_R, \vec{p}) ~ e^{2\pi i  \tau E_L - 2 \pi i \bar{ \tau} E_R + 2\pi i \mu_i p^i}\,,
\ee
where $E_L, E_R, p^i$ are the eigenvalues of $L_0, \bar{L}_0, P^i$ respectively.
For small $\tau$, \eqref{mod transfn} implies that
\be
Z(\tau, \bar{\tau}, \vec{\mu})\approx e^{- \frac{2\pi i \mu^2}{ \tau}} e^{-{2\pi i E_L^v\over \tau }+{2\pi i  E_R^v\over \bar{\tau}}+{2\pi i \mu_i p^i_v\over \tau}}\,.
\ee
Then, inverting \eqref{Z vs d}, we obtain the following expression for the density of states:
\be\label{rhoCFT}
\rho (E_L, E_R, \vec{p}) \simeq \int d\tau d\bar{\tau} \, d^{n} \mu \,  e^{2\pi i \left(-\frac{ \mu^2}{ \tau}- \frac{E_L^v}{\tau} +  \frac{E_R^v}{\bar{\tau}}-E_L \tau + E_R \bar\tau - \mu_i p^i  \right)}\,,
\ee
where we have assumed that the vacuum is electrically neutral, $p^i_v=0$. 
This integral may be evaluated by saddle point methods. The integrand reaches an extremum at
\be \label{saddlepoint}
\tau_0 =  \sqrt{\frac{4 E_L^v}{  4 E_L - \mathcal{P}^2  }} \,, \qquad
\bar{\tau}_0 = - \sqrt{\frac{E_R^v}{E_R }} \,, \qquad	
\mu_{0i} = - k_{ij} p^j   \sqrt{  \frac{E_L^v}{ 4 E_L - \mathcal{P}^2  }}  \,,
\ee
where the matrix  $k_{ij}$ is the inverse of $k^{ij}$ and $ \mathcal{P}^2 \equiv  p^i p^j  k _{ij}$. The leading contribution to the entropy is obtained by evaluating (\ref{rhoCFT}) at the saddle (\ref{saddlepoint}). This gives
\be\label{Sleading}
S = \log \rho_0  = 2\pi \sqrt{ -E_L^v  \left(4 E_L - \mathcal{P}^2 \right)} + 2\pi \sqrt{ -E_R^v \left(4 E_R \right)} \,. 
\ee
Putting
\be
E_L^v=E_R^v=-{c\over 24}\,, \quad E_L-{\mathcal{P}^2\over 4}={\pi^2\over 6}c\,T_L^2\,,\quad E_R={\pi^2\over 6}c\,T_R^2\,,
\ee
we have 
\be\label{Cardy-leading-canonical}
S={\pi^2\over 3}c\,T_L+{\pi^2\over 3}c\,T_R\,.
\ee
The analysis of \cite{Guica:2010ej, Compere:2014bia} yields $c = 6 J_{L  \, \textrm{ext}}$ and using the values for $T_L,T_R$ obtained in (\ref{Tnearext}), we see that \eqref{Cardy-leading-canonical} matches the near-extremal Bekenstein-Hawking entropy  (\ref{BHEntropyNearExt}) to linear order in  $\kappa$.
This extends the match of \cite{Guica:2010ej} from the extremal to the near-extremal regime. 

The logarithmic correction $\Delta S$ to the leading entropy (\ref{Sleading}) is generated by Gaussian fluctuations of the density of states (\ref{rhoCFT}) about the saddle  (\ref{saddlepoint}):
\be\label{DeltaSasDetACFT}
\Delta S = -\frac{1}{2} \log {\det {\cal A} \over (2\pi)^{n+2}}\,,
\ee
where ${\cal A}$ is the determinant of the matrix of second derivatives of the exponent in the integrand of (\ref{rhoCFT}) with respect to $\tau$, $\mu_i$, $\bar{\tau}$. We find
\be \label{detA}
\det {\cal A}   = \frac{(2\pi)^{n+2}}{16} \,  \left(- E_L^v \right)^{-\frac{n + 1}{2}} \,  (4 E_L- \mathcal{P}^2 )^\frac{n+3}{2}    \,     \left(- E_R^v \right)^{-\frac{1}{2}}   \, (4 E_R)^\frac{3}{2} \, \det k^{ij} \,.
\ee
We now fix $n=2$ for the left moving $SU(2) \times U(1)$ current algebra corresponding to $SU(2)$ rotations and the gauge field. The $SU(2) \times U(1)$ charges are $p^1=0$ (because $J_R =0$) and $p^2\propto Q_{\textrm{ext}}$. The $U(1)$ Kac-Moody level $k^{22} \equiv k_Q$ is given in Appendix \ref{App ASG}, the $SU(2)$ level is $k^{11} \equiv k_J \propto c$ \cite{Maldacena:1997ih}, and $k^{12} = k^{21} = 0$. Taking into account (\ref{BHchargesExt}), we thus have the following scalings,
\begin{equation} \label{scalingsCFT}
E_L^v \,, E_R^v  \,, E_L-\mathcal{P}^2/4  \,, E_R  \,  \sim a^3 \,,  \quad k_Q\sim a \,, \quad k_J \sim a^3 \; .
\end{equation}
Bringing (\ref{scalingsCFT}) to (\ref{DeltaSasDetACFT}, \ref{detA}), we  obtain
\be\label{DeltaSmicro}
\Delta S =  -5 \log a  \, .
\ee

\subsubsection{$\widehat{U(1)}_R  \times Vir_L$} 
In this subsection we consider a warped CFT, in which the global symmetries are enhanced as
\be\label{enhancement3}
SL(2,R)_R \times U(1)_L \rightarrow \widehat{U(1)}_R  \times Vir_L\,.
\ee
Here $\widehat{U(1)}_R$ is a left moving Kac-Moody algebra whose zero mode $\tilde{R}_0$ generates the right sector time translations in AdS$_2$ and $Vir_L$ is a left moving Virasoro algebra whose zero mode $\tilde{L}_0$ generates the left sector $U(1)_L$ rotational isometry.
The symmetry algebra of our warped CFT is
\begin{eqnarray*}\label{wcftsymmetry}
\left[\tilde{L}_m,\tilde{L}_n\right] &=& (m-n)\tilde{L}_{m+n} + \frac{c}{12}(m^3 - m)\delta_{m+n}\,, \\
\left[\tilde{R}_m,\tilde{R}_n\right] &=& \frac{k_R}{2}m\delta_{m+n}\,,\quad \left[\tilde{L}_m,\tilde{R}_n\right] = -n\tilde{R}_{m+n} \,,
\end{eqnarray*}
where $\tilde{L}_m$ and $\tilde{R}_m$ are the Virasoro and Kac-Moody generators respectively.
Putting the theory on a circle along $\psi$, the partition function  at inverse temperature $\beta$ and angular potential $\theta$ is given by $Z(\beta,\theta)=\textrm{Tr}\,e^{-\beta\tilde{R}_0+i\theta\tilde{L}_0}$. On the other hand, in \cite{Detournay:2012pc} it was shown that by redefining the charges as
\be\label{newcharges}
L_n = \tilde{L}_n - \frac{2}{k_R}\tilde{R}_0 \tilde{R}_n + \frac{1}{k_R} \tilde{R}_0^2 \delta_n\,, \quad R_n = \frac{2}{k_R} \tilde{R}_0 \tilde{R}_n - \frac{1}{k_R} \tilde{R}_0^2 \delta_n\,,
\ee
and putting the theory on the same circle but in the different ensemble\footnote{A change of ensemble may result in different logarithmic corrections to the entropy. However, as explained in Appendix \ref{App changeofensemble}, the change of ensemble corresponding to   the charge redefinitions \eqref{newcharges} here does not imply any change in the logarithmic correction to the entropy.} 
\be
Z(\tau,\bar{\tau})=\textrm{Tr}\,e^{2\pi i \tau L_0-2\pi i\bar{\tau}R_0}\,,
\ee
the partition function obeys the usual CFT modular invariance:
\be
Z(\tau,\bar{\tau})=Z(-1/\tau,-1/\bar{\tau})\,.
\ee
Assuming
\be\label{tau vs beta wcft}
4\pi \tau=\beta_L-\beta_R+i(\beta_L+\beta_R)
\ee
and $4\pi \bar{\tau}=\beta_L-\beta_R-i(\beta_L+\beta_R)$ 
we may then proceed as in the previous section replacing $\bar{L}_0$ with $R_0$ everywhere starting from equation \eqref{Z with chem pot} onwards. We thus arrive at the same results for the leading entropy and its logarithmic correction.

It should be noted that the enhancement \eqref{enhancement3} is somewhat unusual in the context of warped AdS$_3$ \cite{Azeyanagi:2012zd}. A third more natural enhancement $SL(2,R)_R \times U(1)_L \rightarrow Vir_R \times \widehat{U(1)}_L$ in that context is also possible \cite{Compere:2014bia}. However, this case may not be treated as \eqref{enhancement3} above because the arguments of \cite{Detournay:2012pc} do not  apply to the case when the identification in the bulk (along $\psi$) is precisely anti-aligned with the action of $L_0$ (along $t$). It is an important outstanding problem in Kerr/CFT to generalize the arguments of \cite{Detournay:2012pc} to accomodate this case.

\subsection{Match of the macroscopic and microscopic computations}\label{match}

We have already exhibited the match, in the near-extremal regime,  of the bulk and microscopic results for the leading term of the entropy of the five-dimensional Kerr-Newman black hole under consideration: the Cardy formula (\ref{Cardy-leading-canonical}) reproduces the near-extremal Bekenstein-Hawking  entropy (\ref{BHEntropyNearExt}).

We will now show that the logarithmic corrections also agree. In order to furnish a sensible comparison, one must ensure that both results are given in the same ensemble. This is not the case for the macroscopic, (\ref{DeltaSmacro}), and microscopic, (\ref{DeltaSmicro}), results given above. The former assumes the entropy to be a function of the energy $Q[\partial_{\hat{t}}]$ conjugate to the asymptotic time which features in the full black hole solution (\ref{5dmet}), while the latter is instead a function of the energy $Q[\partial_{t}]$ conjugate to the near horizon time which appears in (\ref{5dmetnearext}). The transformation between the macroscopic and microscopic density of states requires a Jacobian factor (Appendix \ref{App changeofensemble}),
\be
\rho_{\textrm{bulk}} = \frac{\delta Q[\partial_{t}]}{\delta Q[\partial_{\hat{t}}]} \, \rho \; .
\ee
Now, from the change of coordinates (\ref{Coordchange}) and the expression for the extremal angular velocity in (\ref{BHangvelext}), we see that this Jacobian scales like
\be\label{Jacobian match}
\frac{\delta Q[\partial_{t}]}{\delta Q[\partial_{\hat{t}}]} \sim a \, .
\ee
Thus
\be
\Delta S_{\textrm{bulk}} = \Delta S + \log a \; , 
\ee
which indeed is satisfied by (\ref{DeltaSmacro}) and (\ref{DeltaSmicro}).

\section*{Acknowledgements}

We thank Monica Guica, Thomas Hartman, Maria J. Rodriguez and Ashoke Sen for useful conversations. AP and AS are supported in part by DOE grant DE-FG02-91ER40654. APP is supported by NSF grant PHY-1504541. OV is supported by the Marie Curie fellowship PIOF-GA-2012-328798.

\appendix
\section{Computation of $k_Q$}\label{App ASG}

In this appendix we compute the level $k_Q$ of the $U(1)$ Kac-Moody algebra associated with the gauge field $A_{\mu}$. We do not perform a full asymptotic symmetry group analysis here. We expect that with appropriate boundary conditions on the gauge field this Kac-Moody is consistent with the rest of the asymptotic symmetries used in Section 3. Here we are particularly interested in deriving the scaling of the level $k_Q$ with $a$.

Thus we assume the $U(1)$ current algebra is generated by
\be
\Lambda_\eta=\eta(\tilde y)\,,
\ee
where $\tilde{y} = \pi T_L \, \psi$.
In modes, the generators
\be
p_n=-2\pi T_L \,e^{{-in\tilde y/ (2\pi T_L)}}\,,
\ee
satisfy the algebra
\be
\left[p_m,p_n\right]=0\,.
\ee
Using the formulas in \cite{Compere:2009dp}, one can compute the central extension in the corresponding Dirac bracket algebra. We find:
\begin{eqnarray}
\left\{Q_{p_m},Q_{p_n}\right\}&=&-im\, 24\pi^2 T_L^2 a e^\delta\tanh 2\delta\,\delta_{m+n}\,.
\end{eqnarray}
The central extension comes entirely from the Chern-Simons term in the action (\ref{5daction}).
Passing to the commutators $\{\,,\}\to -i[\,,]$ we obtain the current algebra,
\be\label{Virasoro-Kac-Moody}
\left[P_m,P_n\right]={k_Q\over 2} m \delta_{m+n}\,,
\ee
with level given by
\be \label{centralChargeandLevel}
k_Q =12\,(2\pi T_L)^2\,ae^\delta\tanh 2\delta\,.
\ee

\section{Change of ensemble}\label{App changeofensemble} 

Under a charge redefinition, $\vec{q}=\vec{q}\,(\vec{q\,}')$, the density of states, $\rho(\vec{q})$, transforms with the appropriate Jacobian factor as
\be
\rho'(\vec{q\,}')=\frac{\partial(q_1,q_2,\ldots)}{\partial(q_1',q_2',\ldots)}\rho(\vec{q})\,.
\ee
The leading piece of the entropy $S=\log\rho$ typically scales like $a^{D-2}$ for large $q\sim a$ and is therefore independent of the change of ensemble. However, the logarithmic correction, which scales like $\log a$, often picks up contributions from the Jacobian factor above. We have seen this explicitly in section \ref{match} where the Jacobian \eqref{Jacobian match} scales with $a$.

Another instance of a change of ensemble was mentioned in relation to the charge redefinitions in \eqref{newcharges}. In this case the Jacobian is
\be
\frac{\partial(L_0,R_0)}{\partial(\tilde{L}_0,\tilde{R}_0)}=2{\tilde{R}_0\over k_R}=2\sqrt{R_0\over k_R}\,.
\ee
However, $k_R\propto c\sim a^3$ \cite{Compere:2014bia} and $R_0\sim a^3$ so  in this instance the Jacobian does not scale with $a$ and therefore the logarithmic correction to the entropy is left intact by this particular change of ensemble.


\begin{thebibliography}{99}
	
\bibitem{Banerjee:2010qc} 
S.~Banerjee, R.~K.~Gupta and A.~Sen,
``Logarithmic Corrections to Extremal Black Hole Entropy from Quantum Entropy Function,''
JHEP {\bf 1103}, 147 (2011)
[arXiv:1005.3044 [hep-th]].

\bibitem{Banerjee:2011jp} 
S.~Banerjee, R.~K.~Gupta, I.~Mandal and A.~Sen,
``Logarithmic Corrections to N=4 and N=8 Black Hole Entropy: A One Loop Test of Quantum Gravity,''
JHEP {\bf 1111}, 143 (2011)
[arXiv:1106.0080 [hep-th]].

\bibitem{Sen:2011ba} 
A.~Sen,
``Logarithmic Corrections to N=2 Black Hole Entropy: An Infrared Window into the Microstates,''
Gen.\ Rel.\ Grav.\  {\bf 44}, no. 5, 1207 (2012)
[arXiv:1108.3842 [hep-th]].

\bibitem{Sen:2012cj}
A.~Sen,
``Logarithmic Corrections to Rotating Extremal Black Hole Entropy in Four and Five Dimensions,''
Gen.\ Rel.\ Grav.\  {\bf 44} (2012) 1947
[arXiv:1109.3706 [hep-th]].

\bibitem{Bhattacharyya:2012wz} 
S.~Bhattacharyya, B.~Panda and A.~Sen,
``Heat Kernel Expansion and Extremal Kerr-Newmann Black Hole Entropy in Einstein-Maxwell Theory,''
JHEP {\bf 1208}, 084 (2012)
[arXiv:1204.4061 [hep-th]].

\bibitem{Sen:2012dw}
A.~Sen,
``Logarithmic Corrections to Schwarzschild and Other Non-extremal Black Hole Entropy in Different Dimensions,''
JHEP {\bf 1304} (2013) 156
[arXiv:1205.0971 [hep-th]].

\bibitem{Sen:2014aja} 
A.~Sen,
``Microscopic and Macroscopic Entropy of Extremal Black Holes in String Theory,''
Gen.\ Rel.\ Grav.\  {\bf 46}, 1711 (2014)
[arXiv:1402.0109 [hep-th]].

\bibitem{Guica:2008mu}
M.~Guica, T.~Hartman, W.~Song and A.~Strominger,
``The Kerr/CFT Correspondence,''
Phys.\ Rev.\ D {\bf 80} (2009) 124008
[arXiv:0809.4266 [hep-th]].

\bibitem{Bredberg:2009pv} 
I.~Bredberg, T.~Hartman, W.~Song and A.~Strominger,
``Black Hole Superradiance From Kerr/CFT,''
JHEP {\bf 1004}, 019 (2010)
[arXiv:0907.3477 [hep-th]].

\bibitem{Bredberg:2011hp}
  I.~Bredberg, C.~Keeler, V.~Lysov and A.~Strominger,
  ``Cargese Lectures on the Kerr/CFT Correspondence,''
  Nucl.\ Phys.\ Proc.\ Suppl.\  {\bf 216} (2011) 194
  [arXiv:1103.2355 [hep-th]].
  
\bibitem{Compere:2012jk} 
G.~Compere,
``The Kerr/CFT correspondence and its extensions: a comprehensive review,''
Living Rev.\ Rel.\  {\bf 15}, 11 (2012)
[arXiv:1203.3561 [hep-th]].

\bibitem{Guica:2010ej}
M.~Guica and A.~Strominger,
``Microscopic Realization of the Kerr/CFT Correspondence,''
JHEP {\bf 1102} (2011) 010
[arXiv:1009.5039 [hep-th]].

\bibitem{Song:2011sr} 
W.~Song and A.~Strominger,
``Warped AdS3/Dipole-CFT Duality,''
JHEP {\bf 1205}, 120 (2012)
[arXiv:1109.0544 [hep-th]].

\bibitem{Hofman:2011zj} 
D.~M.~Hofman and A.~Strominger,
``Chiral Scale and Conformal Invariance in 2D Quantum Field Theory,''
Phys.\ Rev.\ Lett.\  {\bf 107}, 161601 (2011)
[arXiv:1107.2917 [hep-th]].

\bibitem{ElShowk:2011cm} 
S.~El-Showk and M.~Guica,
``Kerr/CFT, dipole theories and nonrelativistic CFTs,''
JHEP {\bf 1212}, 009 (2012)
[arXiv:1108.6091 [hep-th]].

\bibitem{Bena:2012wc} 
I.~Bena, M.~Guica and W.~Song,
``Un-twisting the NHEK with spectral flows,''
JHEP {\bf 1303}, 028 (2013)
[arXiv:1203.4227 [hep-th]].


\bibitem{Azeyanagi:2012zd} 
T.~Azeyanagi, D.~M.~Hofman, W.~Song and A.~Strominger,
``The Spectrum of Strings on Warped $AdS_3 \times S^3$,''
JHEP {\bf 1304}, 078 (2013)
[arXiv:1207.5050 [hep-th]].


\bibitem{Detournay:2012pc} 
S.~Detournay, T.~Hartman and D.~M.~Hofman,
``Warped Conformal Field Theory,''
Phys.\ Rev.\ D {\bf 86}, 124018 (2012)
[arXiv:1210.0539 [hep-th]].


\bibitem{Detournay:2012dz} 
S.~Detournay and M.~Guica,
``Stringy Schroedinger truncations,''
JHEP {\bf 1308}, 121 (2013)
[arXiv:1212.6792 [hep-th]].


\bibitem{Compere:2014bia} 
G.~Compere, M.~Guica and M.~J.~Rodriguez,
``Two Virasoro symmetries in stringy warped AdS$_{3}$,''
JHEP {\bf 1412}, 012 (2014)
[arXiv:1407.7871 [hep-th]].



\bibitem{Solodukhin:1994yz} 
S.~N.~Solodukhin,
``The Conical singularity and quantum corrections to entropy of black hole,''
Phys.\ Rev.\ D {\bf 51}, 609 (1995)
[hep-th/9407001].

\bibitem{Solodukhin:1994st} 
S.~N.~Solodukhin,
``On 'Nongeometric' contribution to the entropy of black hole due to quantum corrections,''
Phys.\ Rev.\ D {\bf 51}, 618 (1995)
[hep-th/9408068].

\bibitem{Fursaev:1994te} 
D.~V.~Fursaev,
``Temperature and entropy of a quantum black hole and conformal anomaly,''
Phys.\ Rev.\ D {\bf 51}, 5352 (1995)
[hep-th/9412161].

\bibitem{Mann:1997hm} 
R.~B.~Mann and S.~N.~Solodukhin,
``Universality of quantum entropy for extreme black holes,''
Nucl.\ Phys.\ B {\bf 523}, 293 (1998)
[hep-th/9709064].

\bibitem{Carlip:2000nv} 
S.~Carlip,
``Logarithmic corrections to black hole entropy from the Cardy formula,''
Class.\ Quant.\ Grav.\  {\bf 17}, 4175 (2000)
[gr-qc/0005017].

\bibitem{Ferrara:2011qf} 
S.~Ferrara and A.~Marrani,
``Generalized Mirror Symmetry and Quantum Black Hole Entropy,''
Phys.\ Lett.\ B {\bf 707}, 173 (2012)
[arXiv:1109.0444 [hep-th]].

\bibitem{Keeler:2014bra}
C.~Keeler, F.~Larsen and P.~Lisbao,
``Logarithmic Corrections to $N \geq 2$ Black Hole Entropy,''
Phys.\ Rev.\ D {\bf 90} (2014) no.4,  043011
[arXiv:1404.1379 [hep-th]].

\bibitem{Chowdhury:2014lza}
  A.~Chowdhury, R.~K.~Gupta, S.~Lal, M.~Shyani and S.~Thakur,
  ``Logarithmic Corrections to Twisted Indices from the Quantum Entropy Function,''
  JHEP {\bf 1411} (2014) 002
  [arXiv:1404.6363 [hep-th]].

\bibitem{Larsen:2014bqa} 
F.~Larsen and P.~Lisbao,
``Quantum Corrections to Supergravity on AdS$_2\times S^2$,''
Phys.\ Rev.\ D {\bf 91}, no. 8, 084056 (2015)
[arXiv:1411.7423 [hep-th]].

  
\bibitem{Charles:2015eha}
A.~M.~Charles and F.~Larsen,
``Universal corrections to non-extremal black hole entropy in $ \mathcal{N}\ge 2 $ supergravity,''
JHEP {\bf 1506} (2015) 200
[arXiv:1505.01156 [hep-th]].


\bibitem{Belin:2016knb}
  A.~Belin, A.~Castro, J.~Gomes and C.~A.~Keller,
  ``Siegel Modular Forms and Black Hole Entropy,''
  arXiv:1611.04588 [hep-th].
  
\bibitem{Maldacena:1997ih} 
J.~M.~Maldacena and A.~Strominger,
``Universal low-energy dynamics for rotating black holes,''
Phys.\ Rev.\ D {\bf 56}, 4975 (1997)
[hep-th/9702015].

\bibitem{Compere:2009dp}
G.~Compere, K.~Murata and T.~Nishioka,
``Central Charges in Extreme Black Hole/CFT Correspondence,''
JHEP {\bf 0905} (2009) 077
[arXiv:0902.1001 [hep-th]].




\end{thebibliography}
\end{document}